\documentclass[10pt]{article}   
\pdfoutput=1
\usepackage{jheppub}
\usepackage{amsmath,amssymb,amsfonts,mathbbol,graphicx,slashed,color,amsthm, mathtools, upgreek, enumerate, tensor}

\usepackage{subcaption}
\usepackage{setspace}
\usepackage[export]{adjustbox}
\usepackage{arydshln}
\usepackage[dvipsnames]{xcolor}
\usepackage{physics}
\usepackage{hyperref}
\usepackage{comment}
\graphicspath{{pics/}}
\usepackage{breqn}

\usepackage{comment}

\usepackage{algorithm}
\usepackage{algpseudocode}

\usepackage{color}
\definecolor{darkgreen}{RGB}{0,120,0}

\allowdisplaybreaks

\colorlet{darkblue}{blue!70!black}

\colorlet{darkgreen}{green!50!black}
\colorlet{darkred}{red!50!black}

\def\bea{\begin{eqnarray}}
\def\eea{\end{eqnarray}}
\def\be{\begin{equation}}
\def\ee{\end{equation}}

\title{Quantum stress-energy at timelike boundaries: testing a new beyond-$\Lambda$CDM parameter with cosmological data
}

\author[a,b,c,d]{Oliver H.\,E. Philcox,}
\author[c, d]{Eva Silverstein,}
\author[e]{Gonzalo Torroba}
\affiliation[a]{Simons Society of Fellows, Simons Foundation, New York, NY 10010, USA}
\affiliation[b]{Center for Theoretical Physics, Columbia University, New York, NY 10027, USA}
\affiliation[c]{Leinweber Institute for Theoretical Physics at Stanford, 382 Via Pueblo, Stanford, CA 94305, USA}
\affiliation[d]{Kavli Institute for Particle Astrophysics and Cosmology, 382 Via Pueblo, Stanford, CA 94305, USA}
\affiliation[e]{Centro At\'omico Bariloche, CONICET, and Inst. Balseiro, Bariloche, RN, Argentina}
 
 \vspace{5mm}

\vspace{1cm}

\abstract{

\noindent We analyze the basic cosmological effects of a population of timelike boundaries -- a form of nontrivial spacetime topology -- containing a boundary layer of quantum stress energy.  This accumulation of vacuum fluctuations of quantum fields can be consistently negative and UV sensitive, providing an additional source of cosmic energy density strong enough to compete with matter and dark energy.  For boundary conditions enabling a solution with fixed comoving boundary size, this effect contributes a qualitatively new term to the Friedmann equation determining the expansion history, scaling like $-1/a$ for scale factor $a$.  It naturally dominates at relatively late times ($a\approx1/2$), while leaving intact well-measured early universe physics such as big bang nucleosynthesis and recombination. For a wide window of parameters, the boundaries can be larger than the Planck length throughout their history, back through the start of inflation at any viable scale.  We analyze CMB and BAO data sets (Planck, ACT, and DESI) allowing for this component, finding a slight preference ($\sim 2\sigma$) and a relaxation of current tensions in the data (including the neutrino mass) in a physical manner. This novel parameter fits into a larger space of physical parameters beyond-$\Lambda$CDM that may serve this role, including negative spatial curvature, which may also be motivated by topological considerations and chaotic dynamics.  Finally, we comment on additional phenomenological prospects for testing for this form of topology in the universe.  

 }

\begin{document}

\maketitle
\parskip=10pt

\section{Introduction}\label{sec:intro}





\noindent Our cosmological expansion history exemplifies the fundamentals of gravitation, which is well understood to arise from curved spacetime geometry interacting with matter stress-energy.  In addition to its geometry, the topology of our spacetime is a basic feature of the universe, which remains unknown.
Mathematically there are many forms of nontrivial topology that may exist, some of which have been constrained at the scale of our observable horizon using cosmological data (see e.g. \cite{Bond:1999tf, Cornish:2003db,COMPACT:2022gbl}). 

In the present work, we initiate a phenomenological study of a distinctive form of topology: a population of localized timelike boundaries of spacetime. Such boundaries are distinct from ordinary matter, excising a region around a timelike trajectory in spacetime while not classically warping the geometry at all.  Quantum stress energy automatically accumulates near such boundaries  and can be negative and large in magnitude, growing inversely with a power of the distance to the boundary, thus scaling like positive powers of the UV cutoff scale \cite{Deutsch:1978sc}. With this incorporated at a reasonable ultraviolet (UV) cutoff scale, we analyze their effect on large-scale cosmic evolution, focusing on the case of boundary conditions consistent with fixed comoving size expansion of spherical boundaries.  This introduces a new component into the Friedmann equation
\begin{equation}\label{eq-friedmann-intro}
 H^2 \equiv   \left(\frac{\dot a}{a}\right)^2 = \hat\rho_{\Lambda\rm CDM} + \frac{{\hat\rho}_{\rm bdy}}{a} -\frac{k}{a^2} = \left(\frac{\hat\rho_{\rm m}}{a^3} + \frac{\hat\rho_{\rm r}}{a^4} + \hat\rho_\Lambda\right)+\frac{{\hat\rho}_{\rm bdy}}{a}   -\frac{k}{a^2}
\end{equation}
where $\hat\rho_{\rm m}$, $\hat\rho_{\rm r}$, $\hat\rho_{\rm bdy}$ and $\hat\rho_{\Lambda}$
refer to (scaled) matter, radiation, boundary, and dark energy densities respectively. The model parameters including the boundary size at the start of inflation, the density, and the UV cutoff scale -- which together determine $\hat\rho_{\rm bdy}$ -- must be  accessible at effective field theory (EFT)  scales throughout the history and consistent with established big bang nucleosynthesis (BBN) and recombination physics. 
Finding this effect to be observationally accessible and indeed viable in a wide window of parameters, we constrain it with up to date cosmic microwave background (CMB) and baryon acoustic oscillations (BAO) measurements including Planck, ACT and DESI. This analysis reveals a slight preference for such a boundary component with  {$\hat\rho_{\rm bdy}<0$}, which somewhat relaxes current tensions in the data \citep[e.g.,][]{DESI:2024mwx,DESI:2025zgx,DESI:2025gwf,DESI:2025wyn,DESI:2025fii,Malekjani:2024bgi,DESI:2024kob,Shlivko:2024llw,Cortes:2025joz,Wolf:2025jed,Colgain:2024mtg}.  

This is a novel example of a physical way to relax such tensions, analogous to negative curvature ($k<0$ in \eqref{eq-friedmann-intro}), recently demonstrated in \cite{DESI:2025zgx, Chen:2025mlf}. Interestingly, negative curvature may also be motivated by its connection to generic topology \cite{thurston1997three, 2008arXiv0809.4881M}.  Beyond the motivation of their genericity, both types of topology introduce an interesting dynamical feature.  They promote chaos in the dynamics of early universe fields, helping to homogenize the system independently of inflation.  This effect was discussed for the case of hyperbolic spatial geometry without boundaries in e.g. \cite{Cornish:1996st, Linde:2004nz}.  The topology in our scenario -- spherical boundaries -- corresponds to another classic case of chaotic dynamics:  billiards problems \cite{chernov2006chaotic}.  In both cases, particle trajectories separate and efficiently explore the space, homogenizing the system.      

In related work, as explained in \cite{Craig:2024tky, Green:2024xbb, Lynch:2025ine, Graham:2025dfg}, the tensions in current data can be formally relaxed by allowing the neutrino mass to take negative values, which could for example indicate physical explanations involving clustering.  The DESI collaboration has explored various models and parameterizations of the data, including regimes with apparent violation of the null energy condition \citep{DESI:2025wyn,DESI:2025fii,DESI:2024kob}.  The present scenario allows for a limited amount of negative quantum stress-energy, and thus a reduction to the Universe's expansion rate, contributing naturally at late times.

Additional tests of boundaries within our observationallly accessible patch of the universe can be formulated for a range of parameters.  For sufficiently large boundaries, microwave background radiation reflects off the boundaries, changing the last scattering geometry sourcing localized distortions analogous to those from massive fields \citep[e.g.,][]{Flauger:2016idt,Munchmeyer:2019wlh,Philcox:2024jpd}.  We leave a detailed analysis of this for future work.\footnote{Work in progress with Colin Hill}  

As we will review briefly in the next section, the consistency of gravity in the presence of such boundaries is a subtle question of ongoing research interest in physics and mathematics \cite{Anderson:2003un, An:2021fcq, Andrade:2015gja, Marolf:2012dr, Witten:2018lgb, Anninos:2023epi, Anninos:2024wpy, Liu:2024ymn, Liu:2025xij}, with some cases \cite{Fournodavlos:2021eye, Fournodavlos:2020wde, An:2025gvr} reportedly proved consistent at the nonlinear level. This is important to sort out, for the simple reason that such topologies are generic if they exist. 
Another strong theoretical motivation is the recent realization that finite timelike boundaries with frozen, spherical, boundary geometry (Dirichlet boundary conditions) enable a non-gravitational holographic formulation of the positive cosmological constant similar to the AdS/CFT correspondence, revealing familiar thermodynamic relations \cite{Banihashemi:2022htw, Banihashemi:2022jys} and a microstate count of the refined \cite{Anninos:2020hfj} Gibbons-Hawking \cite{Gibbons:1977mu} entropy (see \cite{Coleman:2021nor, Batra:2024kjl} for $dS_3$ and \cite{Silverstein:2024xnr} for a generalization to $dS_4$ building from \cite{Hartman:2018tkw}).  Quantum effects have played some role in this setting as well \cite{Levine_2023, Batra:2024qju}, and interesting steps have been made toward a string-theoretic understanding \cite{Ahmadain:2024hgd, Ahmadain:2024uyo, Silverstein:2022dfj}.   It is known that Dirichlet boundary conditions for gravity are sometimes inconsistent \cite{Anderson:2003un}, but specific regimes including a spherical boundary geometry evades the known problems \cite{Anninos:2023epi, An:2025gvr}.  The dynamics of timelike boundaries includes vaious linearized instabilities \cite{Andrade:2015gja, Anninos:2024wpy}, which are not always robust against perturbative changes to the boundary conditions \cite{Anninos:2023epi}.  Such instabilities may provide new types of signatures, though here we will focus on a scenario in which these are not dominant.   

This paper is organized as follows. In \S\ref{sec-tlike-intro-theory-classical-quantum}, we introduce the relevant classical and quantum physics of a population of timelike boundaries, including various subtleties.  In the remainder of the paper we treat this possibility as a phenomenological scenario.    In \S\ref{sec-basic-scales-viability} we then analyze the scales in the problem to verify that this scenario can happen under control and at a significant enough amplitude to affect cosmological evolution at a detectable level.  We then exhibit in \S\ref{sec-preserve-inflation-BBN-recombination} a wide window of parameters for which the boundaries component does not interfere with primordial curvature fluctuations, BBN and recombination.  With this in place, in \S\ref{sec-data-analysis} we present an analysis of the probability distribution for $\hat\rho_{\rm bdy}$ in concert with existing cosmological parameters, showing its potential viability in easing the current DESI tensions. {We conclude in \S\ref{sec-discussion}, including a discussion of other physical signatures generated by our boundary component}.

\section{Classical and quantum physics of timelike boundaries and their subtleties}\label{sec-tlike-intro-theory-classical-quantum}

Establishing the consistency of timelike boundaries of spacetime is nontrivial. 
This includes, but goes well beyond, the need for a well posed variational problem.  That in itself is straightforward to satisfy by incorporating an appropriate boundary term in the action.  For example, for a Dirichlet condition where we fix the intrinsic geometry of the boundary, the appropriate action is 
\bea\label{eq-action-Lorentzian}
S &=& \frac{1}{16\pi G} \int_M d^{4}x\,\sqrt{-g} \left(R^{(4)} + \mathrm{matter} \right) +\frac{1}{8\pi G} \int_{\partial M} d^3x\,\sqrt{-h}\left( K + \mathrm{counterterms} \right)  
\eea
for the bulk ($M$) plus the boundary ($\partial M$), with $K$ the trace of the extrinsic curvature of the boundary.

Beyond this, consistency requires the existence of solutions respecting the boundary condition, and that they be determined uniquely from initial conditions within the non-excised part of the spacetime.
To see how this Cauchy problem can fail, consider a Dirichlet condition fixing the intrinsic geometry of the boundary, so that gravity is not dynamical on the boundary itself.  That means that modes which are pure gauge in the absence of the boundary can become physical on it.  There exist such modes in the linear perturbation spectrum around a flat wall in Minkowski spacetime which ruin the uniqueness of the evolution.  {For spherical boundaries, we can organize the modes in terms of their angular dependence along the directions of the sphere and radial dependence transverse to it.}   The problematic modes disappear for a spherical boundary for any finite angular momentum (see \cite{Anninos:2023epi}, \S5.5 for an accessible account of this).  A nonlinear, early time proof of consistency in the regime of interest for our application has been reported recently in \cite{An:2025gvr}.  Early work \cite{Andrade:2015gja} derived a spectrum of growing modes at the linearized level in this problem, although it is interesting to note that \cite{Anninos:2023epi} illustrated sensitivity of growing modes to small perturbations of the boundary conditions.      

The general relativistic analysis alone is not applicable for arbitrarily large angular momentum $\ell$, i.e. arbitrarily short scales.  Quantum gravity effects become strong at the Planck scale, if not before -- in perturbative string theory there are effects that soften amplitudes at a longer distance scale (that of the string tension). Arbitrarily large angular momentum can be of interest for the question of the consistency of classical GR per se, as discussed in a recent analysis of a different type of boundary condition \cite{Anderson:2006lqb} in \cite{Liu:2025xij} where an instability derived in \cite{Anninos:2023epi, Anninos:2024wpy} leads to ill-posedness of the Cauchy problem in a particular large-$\ell$ limit. As just mentioned, this does not directly apply to the quantum problem, and even at the classical level it was shown in \cite{Anninos:2023epi} (\S5.4) how one can remove unstable modes by perturbing the boundary conditions.   

In this work, we do not solve the general quantum gravity problem of timelike boundaries, and instead treat the problem phenomenologically.  We consider any boundary condition which yields fixed comoving size boundaries to good approximation.  Our main consideration is the effect of quantum fields in this setting on the cosmic evolution.  

Quantum field theory in a spacetime with boundaries exhibits strong quantum effects \cite{Deutsch:1978sc}; see \cite{Vassilevich:2003xt} for a review and references.  In particular, the expectation value of the stress-energy tensor diverges near the boundary (which we here assume to be a wall for illustration),
\be
\langle T_{\mu\nu} \rangle = \frac{T_{\mu\nu}^{(4)}}{\Delta x_\perp^4}+ \ldots + \frac{T_{\mu\nu}^{(1)}}{\Delta x_\perp}+ \mathcal O(\Delta x_\perp^0)\,,
\ee
where $\Delta x_\perp$ is a short-distance cutoff that sets a minimum distance away from the wall. The first factor $T_{\mu\nu}^{(4)}$ is dimensionless, while the other divergent terms are controlled by local geometric quantities associated to the wall, such as its area and the extrinsic and intrinsic curvatures. 

The short distance dependence of $\langle T_{\mu\nu} \rangle$ exhibits the UV sensitivity of quantum effects near walls. A simple way to understand the origin of these UV effects is to consider a free scalar field with a boundary condition that can be implemented in terms of image sources (such as Dirichlet or Neumann boundary conditions). Then evaluating $\langle T_{\mu\nu} \rangle$ requires taking derivatives of the coincidence limit of the Greens' function in the presence of walls. This produces additional divergences, compared to the result in the absence of walls, corresponding to the coincidence limit of the source and its image \cite{Deutsch:1978sc}. This UV sensitivity can also be understood to arise from the contribution of high frequency modes to the sum of zero point energies. We review this in Appendix \ref{app:zeropoint}, also exhibiting how different boundary conditions can lead to quantum contributions of different sign to the energy density. 

In a UV complete theory of quantum gravity, we expect $1/\Delta x_\perp$ to be of order of the Planck scale or the string scale. A simple example  exhibiting the resulting negative energy density possible in this setting corresponds to a massless scalar field with Dirichlet boundary conditions, which yields an energy density $\langle T_{00} \rangle \sim - 1/\Delta x_\perp^4$ \cite{Deutsch:1978sc}. Similarly, at order $1/\Delta x_\perp^3$, both Dirichlet and Neumann boundary conditions contribute negatively to the energy density outside a spherical wall, with a coefficient that is proportional to the extrinsic curvature of the wall.  

{Although consistently negative, this stress energy is bounded from below by the UV cutoff scale.  Its dynamics are also limited:  for boundary conditions which change smoothly with time, the boundary stress tensor \cite{Brown:1992br} is approximately conserved and the system cannot descend to arbitrarily negative energy.  Altogether, the possibility of negative QFT energy accumulating at a population of timelike boundaries is theoretically viable, though subtle and an active research direction.  In the remainder of this paper we analyze this scenario phenomenologically and observationally},  {seeking to answer whether it can reasonably affect cosmological evolution.}

\section{Can quantum energy contribute detectably to cosmic evolution?}\label{sec-basic-scales-viability}

In this section we make a very basic check of our scenario in which the observable universe contains a collection of $N_{\rm bdy}$ spherical boundaries with negative QFT stress energy as in \S\ref{sec-tlike-intro-theory-classical-quantum}, focusing on the case with fixed comoving size.  We will check that in a reasonable window of scales for our parameters (boundary size, UV cutoff scale $M_{\rm UV}$, number density, and inflationary reheating scale), this component would have always been within the regime of control of effective field theory (EFT) while contributing significantly to cosmic evolution via the Friedmann equation \eqref{eq-friedmann-intro}.   

We prescribe a number density of spherical boundaries in the current universe
\begin{equation}\label{eq-current-n}
    n_{{\rm bdy},0}\sim N_{\rm bdy} H_0^3
\end{equation}
in terms of the current Hubble scale, $H_0$.  
For each boundary, there is a divergent stress energy as we approach the boundary \cite{Deutsch:1978sc}.  As discussed in \S\ref{sec-tlike-intro-theory-classical-quantum}, this divergence is a UV sensitive power law depending on the specifics of the field content.  
With a UV cutoff scale $M_{\rm UV}$, there is a local energy density within a thickness $\sim M_{\rm UV}^{-1}$ near the boundary of order
\begin{equation}\label{eq-cutoff-divergence-powers}
    -{{\cal R}}{M_{\rm UV}^2}, ~~ -{{\cal K}}{M_{\rm UV}^3}, ~~ -M_{\rm UV}^4\,.
\end{equation}
 {where ${\cal R}$ and ${\cal K}$ encode the intrinsic and extrinsic curvatures.} This gives an energy $E_{\rm bdy}$ per boundary of area $A_{\rm bdy}$ of order 
\begin{equation}\label{eq-energy-per-boundary}
-M_{\rm UV}, ~~ -A_{\rm bdy} {\cal K} M_{\rm UV}^2, ~~ -A_{\rm bdy} M_{\rm UV}^3
\end{equation}
where we used that $A_{\rm bdy}{\cal R}\sim 1$. 


To check that this scenario is viable, we consider a number of necessary conditions in the remainder of this section and in \S\ref{sec-preserve-inflation-BBN-recombination}. In particular:
\begin{itemize}
    \item Do the boundaries appreciably affect the Universe's expansion?
    \item Is their size and energy appropriate?
    \item Are they sufficiently homogeneous?
    \item Do they affect the well-constrained physics of inflation, BBN, and recombination?
\end{itemize}

We can determine a number of boundaries $N_{\rm bdy}$ needed to contribute to the energy density at the percent level\footnote{In \S\ref{sec-data-analysis}, we will find a slight preference in the CMB and LSS data for a 10\% contribution from boundaries. This is viable within a similar window of model parameters.} as follows 
\begin{equation}\label{eq-rho-match}
    \rho_{{\rm bdy}, 0}=N_{\rm bdy} E_{\rm bdy} H_0^3 \sim -10^{-2}H_0^2 M_P^2 \quad\Rightarrow\quad N_{\rm bdy} \sim -10^{58}\frac{M_P}{E_{\rm bdy}}
\end{equation}
where we have written the critical density in terms of the Planck mass, noting that $M_P/H_0\sim 10^{60}$.  We note that, consistently with sub-Planckian cut off energy densities, the ratio $M_P/E_{\rm bdy}$ can be $>1$ or $<1$ depending on the case.  In the last two cases of \eqref{eq-energy-per-boundary} we can have $E_{\rm bdy} > M_P$ (since the area can be large), while in the first case $E_{\rm bdy}\le M_P$.  

There is another bound on $N_{\rm bdy}$ from requiring that the number density at the start of inflation is sub-Planckian, that the spheres fit in a non-overlapping way, and that they are themselves larger in size than $M_P^{-1}$:  
\begin{equation}\label{eq-nbstart-small-Planck-condition}
    n_{\rm bdy, start} \ll \frac{1}{r_{\rm bdy, start}^3} \ll M_P^3
\end{equation}
where $r_{\rm bdy, start}\sim A_{\rm bdy, start}^{1/2}$ is the proper sphere radius at the start of inflation
(after which it dilutes and becomes more and more sub-Planckian.)
The densities at reheating and at the start of inflation can be obtained by rescaling the current number density \eqref{eq-current-n}:
\begin{equation}\label{eq-nb-reh}
    n_{\rm bdy, reh}=n_{\rm bdy,0}\frac{a_0^3}{a_{\rm reh}^3}=n_{\rm bdy,0}\frac{T_{\rm reh}^3}{T_{0}^3}
\end{equation}
and
\begin{equation}\label{eq-nb-start}
    n_{\rm bdy,start}=n_{{\rm bdy,0}}\frac{T_{\rm reh}^3}{T_{0}^3}e^{3 N_e}=n_{{\rm bdy,0}}\frac{T_{\rm reh}^6}{T_0^3 T_{\rm GUT}^3} e^{3\times 60} \ll \frac{1}{r_{\rm bdy, start}^3} \ll M_P^3
\end{equation}
with $T_0\sim 10^{-4}\mathrm{eV}=10^{-13}\rm{GeV}$ and $T_{\rm GUT}\sim 10^{15}\rm{GeV}=10^{24}\rm{eV}$.  This yields a constraint on the number $N_{\rm bdy}$ of boundaries
\be\label{eq-Nb-bound}
    N_{\rm bdy} \ll \frac{1}{r_{{\rm bdy, start}}^3 M_P^3}\left(\frac{M_P}{H_0}\right)^3 e^{-3\times 60} \left(\frac{T_{\rm reh}}{10\rm GeV}\right)^{-6} 
    \sim \frac{1}{r_{{\rm bdy, start}}^3 M_P^3}\left(\frac{T_{\rm reh}}{10^{17} \rm GeV}\right)^{-6}\,.
\ee
Since current bounds on primordial gravitational waves, which limit the scale of inflation, ensure $T_{\rm reh} < 10^{15} \mathrm{GeV}$ \citep[e.g.,][]{Lyth:1996im,BICEP:2021xfz}, the last factor on the RHS of this inequality is generally $\gg 1$ as a necessary condition.  We need $r_{\rm bdy} M_P \gg 1$ just by a modest control factor, so our constraint is very easy to satisfy. 

It is interesting to combine the above result with \eqref{eq-rho-match}, giving
\begin{equation}\label{eq-second-condition-Nb-etc}
    \frac{M_P}{E_{\rm bdy}}\sim 10^{-46} \left(\frac{T_{\rm GUT}}{T_{\rm reh}}\right)^6
\end{equation}
If the divergence of the stress-energy at the boundary is only quadratic (the first case in \eqref{eq-cutoff-divergence-powers} and \eqref{eq-energy-per-boundary}), which implies that the LHS of \eqref{eq-second-condition-Nb-etc} is $>1$, we need $T_{\rm reh}/T_{\rm GUT} \ll 10^{-8}$, which is a strong constraint on the reheating temperature.  For the other cases, however, there is more room for higher reheating temperatures; this scenario is consistent with any viable scale of inflation.


For the scenario we are considering in this paper, the most basic requirement is that the set of boundaries be sufficiently homogenous that we can approximate them as such in estimating their effect on the expansion history.  This demands that the typical inter-boundary distance $d_{\rm bdy}$ be much smaller than the size of the observable universe: 
\begin{equation}\label{eq-basic-lower-bound-Nb}
    d_{\rm bdy}H \ll 1\,.
\end{equation} 
 {This simply requires $N_{\rm bdy}\gg 1$, though we will find stronger constraints below by requiring that the recombination history is not strongly impacted by the boundaries.}




\section{Preserving the prehistory: inflationary modes, BBN and recombination}\label{sec-preserve-inflation-BBN-recombination}

In this section we verify that there exists a regime where the density of boundaries does not significantly modify observables derived from the primordial inflationary curvature perturbations and from CMB physics originating at recombination. In this analysis, we can consider two cases for their spatial distribution:  a regular distribution, leading to Bloch wave mode solutions, or a random distribution in which one should assess whether Anderson localization arises\footnote{We thank Daniel Green for raising this question.}.  In both cases, the physics depends on the hierarchy between the relevant wavelengths and the typical distance between boundaries.   We can avoid strong effects by establishing either high density or low density compared to each of the relevant wavelengths.

At high density, the short scale features decouple from the long wavelength physics we observe, although long-time effects such as refraction can be important\footnote{We thank Peter Graham for discussing this effect with us in the context of a different scenario.}.  This is an example of dangerous irrelevance, where the na\"ive energetic decoupling does not screen UV physics from IR effects.  At low density, the boundary effects (scattering) can be subdominant to the standard dynamics.  We will focus on this case below, as it turns out not to  {be significantly constrained by} the inflationary scale.

\subsection{Primordial curvature perturbations}\label{sec-primordial-curvature}

In the case of the primordial perturbations, the wavelengths are exponentially expanding; for a given comoving scale ${k}$ the mode evolves from a short vacuum fluctuation until it freezes out at the inflationary Hubble scale $k/a_f = H_{\rm inf}$.  

Regardless of the time dependence of the proper size of the boundaries, their separation grows $\propto a(t)$.  We aim to work in a regime where the wavelength is long compared to the spacing between boundaries,
\begin{equation}\label{eq-primordial-long-wavelength}
    \frac{k/a}{n_{\rm bdy}^{1/3}} \ll 1
\end{equation}
for some comoving scale of relevance $k$. Since $n_{\rm bdy}\sim a^{-3}$, this is satisfied at all times if it is satisfied at any given time. Here, the relevant scales are those that freeze out early in inflation, reentering the horizon late to produce the CMB+LSS observables in the linear or weakly nonlinear regime. Freezeout happens at 
\begin{equation}
    -k \eta =k/aH \sim 1
\end{equation}
where $\eta\sim -1/aH$ is the conformal time. 

The longest mode $k^{(\ell)}$ we see has a physical wavenumber
$k^{(\ell)}/a_{\rm start}=H_{\rm inf}$, since this mode freezes out right away at the start of inflation.  The shortest mode relevant for quasi-linear CMB + LSS is $k^{(s)}\sim k^{(\ell)}e^{N_{e, \rm CMBLSS}}$ where the number of efoldings accessible in the CMB + LSS is $N_{e, \rm CMBLSS}\sim 10$.  The condition for these modes to all see a nearly homogeneous density of boundaries is thus
\begin{equation}
    \frac{k^{(\ell)}}{a_{\rm start}}e^{N_{e, \rm CMBLSS}}\ll n_{{\rm bdy, start}}^{1/3} \Rightarrow H_{\rm inf} \ll n_{{\rm bdy, start}}^{1/3}e^{-N_{e, \rm CMBLSS}}\,.
\end{equation}
Putting this together with the basic condition for control in the previous section, we have
\begin{equation}
    H_{\rm inf} e^{N_{e, \rm CMBLSS}}\ll n_{{\rm bdy, start}}^{1/3} \ll \frac{1}{r_{{\rm bdy, start}}} \ll M_P\,.
\end{equation}
Given the upper bound of order $H_{\rm inf}<10^{-7}M_P$ from the constraint on the tensor to scalar ratio $r$ \citep{BICEP:2021xfz}, this condition is not very constraining. 
Below in \S\ref{sec-recombination} we will recover the consistency of this regime in our scenario, working with these modes evolved to the time of recombination.

\subsubsection{More detailed substructure}

Although not necessary in our regime with a well controlled homogeneous approximation for the primordial perturbations in the linear regime, it is interesting to consider the residual effects of the density scale.  As discussed above, we can consider either a regular distribution of boundaries or a more random one.

For a regular lattice of boundaries, the mode solutions will involve Bloch wavefunctions.  There is a range of $k$ such that the physical wavelength is of order the separation between boundaries.  In this regime, phenomena such as band gaps may appear.  Such gaps would impact the power spectrum at those scales, which we have ensured here are not visible in the CMB + LSS window of (quasi-)linear perturbations. However, this might lead to residual effects, somewhat analogously to the way that primordial black hole production can enter over a range of short scales.  Being deep in the nonlinear regime, such effects are not immediately constraining but may yield some kind of residual signatures.  

Similar comments apply to the case of a random distribution of boundaries.  With sufficiently strong disorder or in low dimensions, Anderson localization can arise.  
This can be heuristically understood as a phenomenon of constructive interference for paths returning to a local region along with destructive interference of propagation.   The mean free path is 
\begin{equation}\label{eq-mean-free-path}
    \ell_{\rm mfp} = \frac{1}{n_{\rm bdy}\sigma}
\end{equation}
and the corresponding localization transition is estimated  to be \cite{AndersonLoc2010arXiv1005.0915M}
\begin{equation}
    k_{\rm phys}\ell_{\rm mfp} \sim 1
\end{equation}
above which the system conducts rather than localizing. 
For the s-wave cross section of order $r_{\rm bdy}^2$ we have
\begin{equation}
    k_{\rm phys}\ell_{\rm mfp, start} \sim \frac{k_{\rm phys}r_{\rm bdy}}{\pi}\frac{1}{{r_{{\rm bdy, start}}^3} n_{{\rm bdy, start}}} \gg 1\,.
\end{equation}
Both factors here are $\gg 1$ in our scenario already, 
{indicating an absence of Anderson localization in our scenario.}

\subsection{Recombination physics and BBN}\label{sec-recombination}

In this section we impose conditions for our scenario to avoid significantly affecting the physics of recombination. 
Having done so, it will be straightforward to ensure that our scenario is also consistent with big bang nucleosynthesis, and to recover the consistency with inflationary perturbations discussed in \S\ref{sec-primordial-curvature}.

It is informative to compare the scattering of particles (mainly photons, electrons, and protons) from the boundaries to their usual scattering from each other in the plasma around the time of recombination. Since for long wavelengths the density of boundaries would act like a nearly homogeneous medium with a modified index of refraction, a hierarchy between these scattering rates is not be required.  For the phenomenology discussed herein to hold, it would be sufficient to have either the scattering rate itself be small, or the index of refraction from the collective effect of the boundaries be a small effect.  We will analyze both regimes here, finding the former to be favorable for our scenario.  


The scattering rates for each type of interaction are given by 
\begin{equation}
    \Gamma = \sigma n v
\end{equation}
in terms of the cross-section $\sigma$, the number density $n$, and the relative velocity $v$.  For the boundaries, their proper size at recombination $r_{\rm bdy, rec}$ enters into the cross section.  Assuming (as we have been) that we have fixed comoving size of the boundary throughout our history including during inflation, and insisting that the boundaries were larger than Planck scale all the way back to the start of inflation, this size is bounded below by:
\begin{equation}
    r_{\rm bdy, rec} > \frac{1}{M_P}\frac{a_{\rm rec}}{a_{\rm start}}= \frac{1}{M_P} \frac{T_{\rm reh}}{T_{\rm GUT}}\frac{T_{\rm GUT}}{T_{\rm rec}}e^{N_e}\sim 10^{16}{\rm m}\left(\frac{T_{\rm reh}}{T_{\rm GUT}}\right)^2
\end{equation}
where we plugged in $M_P^{-1}\simeq 10^{-35}{\rm m}, T_{\rm GUT}/T_{\rm rec}\sim 10^{15} {\rm GeV}/0.1 {\rm eV} \sim 10^{25}, e^{N_e}=e^{60}(T_{\rm reh}/T_{\rm GUT})$.  Today, the boundaries would be $1000\times$ larger.  
This gives a wide range of possible sizes at recombination depending on the (unknown) scale of inflation; assuming that is at least $10{\rm MeV}$ to be consistent with BBN we have
\begin{equation}\label{eq-range-rbdryrec}
      10^{-18}{\rm m} < r_{\rm bdy, rec} < 10^{16} {\rm m}  ~~~~~~~ {\textrm{for}} ~~~~~~~ 10 \rm{MeV} < T_{\rm reh} < T_{\rm GUT}\,.
\end{equation}
(For $\rm{TeV}$ scale inflation this would be $10^{-8}{\rm m}$, for example.)
The other key scale for the boundaries is their density $n_{\rm bdy}$, equivalently their spacing $d_{\rm bdy}$ or volume fraction $f_{\rm bdy}$
\begin{equation}
    f_{\rm bdy} \sim \frac{\frac{4}{3}\pi r_{\rm bdy}^3}{d_{\rm bdy}^3} \sim \frac{4}{3}\pi r_{\rm bdy}^3 n_{\rm bdy} \,.
\end{equation}
Above we noted general upper and lower bounds for $n_{\rm bdy}$, requiring it to be sufficiently homogeneous to affect the evolution history as in \eqref{eq-friedmann-intro} while maintaining $d_{\rm bdy}>r_{\rm bdy}$ by some margin.

Given the factor of 1000 expansion to the current universe, this would translate into a range of sizes 
\begin{equation}
    10^{-15}{\rm m} < r_{{\rm bdy,} 0}<10^{19}{\rm m}
\end{equation}
The upper end of this range, corresponding to GUT scale inflation, leads to boundaries of the size of the visible part of current galaxies:
\begin{equation}\label{eq-size-galaxies}
    r_{\rm galaxy, 0}\sim 10^{19}{\rm m}, ~~~~~~ d_{\rm galaxy, 0} \sim 10^{22}{\rm m}
\end{equation}
(with the dark matter halos 10 times larger or so).  Depending on how easy these large boundaries would be to see in between galaxies (e.g. in cosmic voids, as they repel matter), we could scale this down easily with the scale of inflation.  


It is essential that our boundaries not disrupt the basic observables coming from recombination.  Relevant scales for the usual Standard model physics include
\begin{eqnarray}
     T_{\rm rec} & \sim 3000{\rm K} \\
     \lambda_{\rm linear, short, rec} & \sim 10^{-2} {\rm Mpc} \sim 10^{20}{\rm m} \\ \label{eq-sm-lambda-linear-short-rec}
      ~~~\lambda_{\gamma, \rm CMB, rec} &\sim 10^{-6}{\rm m} \label{eq-standard-model-wavelengths-rec}\\
     \lambda_{\rm dB, e^-, rec}&\sim 10^{-9}{\rm m}\\
     \lambda_{\rm dB, p^+, rec}&\sim 10^{-10}{\rm m}\\
     d_{\rm atomic} &\sim 10^{-10}{\rm m}  \label{eq-sm-wavelengths-again-1}
\end{eqnarray}\label{eq-sm-wavelengths-again}where $\rm dB$ refers to the de Broglie wavelengths of the proton and electron and $\lambda_{\rm linear, short, rec}$ is the scale of the shortest density perturbations evolved from the primordial curvature perturbation, within the linear regime accessible to CMB and LSS .  

A priori there are two regimes to consider, depending on the relative scales.  

{\bf (i)}  There is a broad regime where the density is sufficient to be homogeneous for the purpose of FRW evolution -- meaning we have many boundaries in a Hubble patch (which is of order the particle horizon) -- while being small enough to not strongly affecting the recombination dynamics.  To see this very simply, we compare the scattering rate off the boundaries to Hubble at recombination.  The scattering among Standard Model particles in the plasma is faster than Hubble, so it is sufficient to bound the scattering off boundaries to be less frequent than Hubble.

Taking for simplicity the regime where the Standard Model wavelengths are also all small compared to $r_{\rm bdy}$, imposing this sufficient condition to not interfere with recombination requires
\begin{equation}\label{eq-Gamma-rec-ineq}
    \Gamma = \sigma n v \sim r_{\rm bdy, rec}^2 n_{\rm bdy, rec} v < H_{\rm rec}
\end{equation}
and since $v\le c$ it is sufficient to impose this inequality with $v$ replaced by $c$ (=1 in natural units), giving
\begin{equation}\label{eq-recwindow}
    H_{\rm rec}^3 \ll n_{\rm bdy, rec} < \frac{H_{\rm rec}}{r_{\rm bdy, rec}^2}
\end{equation}
with the first inequality ensuring that there are many boundaries per Hubble patch.  Since we automatically have $r_{\rm bdy, rec} \ll H_{\rm rec}^{-1} \sim 10^{22}{\rm m}$ from \eqref{eq-range-rbdryrec},
there is a large window available for this, consistently with any scale of inflation.  

Moreover, this window is also consistent with the earlier epochs of inflation and BBN. Given the scale \eqref{eq-sm-lambda-linear-short-rec} of the primordial perturbations evolved to recombination we have $H_{\rm rec}^{-1} \sim 10^2 \lambda_{\rm linear, short, rec}$. Thus it is straightforward to modify the window \eqref{eq-recwindow} slightly to ensure that $\lambda_{\rm linear, short, rec}\gg n_{\rm bdy}^{-1/3}$ as discussed during inflation in \S\ref{sec-primordial-curvature}.  The nuclear scales involved in BBN are shorter than the atomic scales relevant for recombination, so scaling back to that time we would find again that the wavelengths are shorter than the boundary size.  The rate $\Gamma$ on the left hand side of the inequality \eqref{eq-Gamma-rec-ineq} scales up by a factor of $a_{\rm rec}/a_{\rm BBN}$, while the right side (Hubble) scales up approximately by a factor of $(a_{\rm rec}/a_{\rm BBN})^{1/2}$ since it was mostly radiation dominated during the intervening period.  The ratio between the two is $\sim 10^{5/2}$, not many orders of magnitude and hence easily accommodated.  In this regime, we can also look for direct CMB signatures of individual boundaries, which is briefly discussed in \S\ref{sec-discussion}.

{\bf (ii)}  At the other extreme end of parameter space, for a window of very minimal, low scale, inflation, the wavelengths \eqref{eq-standard-model-wavelengths-rec}-\eqref{eq-sm-wavelengths-again-1} can all be longer than the boundary scales $d_{\rm bdy, rec} \gg r_{\rm bdy, rec}$.  In this long wavelength limit, the density of boundaries is effectively homogeneous as seen by all constituents of the Standard model during recombination, which are not energetically sensitive to the details of the short scale features.  

However, for case (ii) we would need to check the integrated effect of the density on CMB photons that travel from last scattering to our detectors.  This could be an example of dangerous irrelevance, where the long propagation time builds up strong enough effects to constrain this, even though energetically the Standard Model particles do not probe the $d_{\rm bdy}$ scale between boundaries.  In particular, the density of boundaries introduces an index of refraction controlled by the volume fraction.  There are various strong bounds on the speed of light, at least relative to the gravitational waves (though in our case, both are similarly affected by the boundaries), and on frequency-dependence in the speed of light.  Indeed, initial estimates suggest that e.g. \cite{Nemiroff:2011fk} could exclude this regime.  Since this regime would heavily constrain the scale of inflation in any case, we do not pursue it further here.

In summary, in the regime (i) defined above, our scenario is consistent with early universe Standard model physics, entering just as a background energy density.  Its effect on the Friedmann equation \eqref{eq-friedmann-intro} dominates at relatively late times, when current data exhibits hints of tensions.  We turn now to analyze the probability distribution for this parameter according to current data. 


\section{Analysis of cosmological parameters in the presence of boundaries}\label{sec-data-analysis}



\subsection{CMB and LSS Phenomenology}
\begin{figure}
    \centering
    \includegraphics[width=0.8\linewidth]{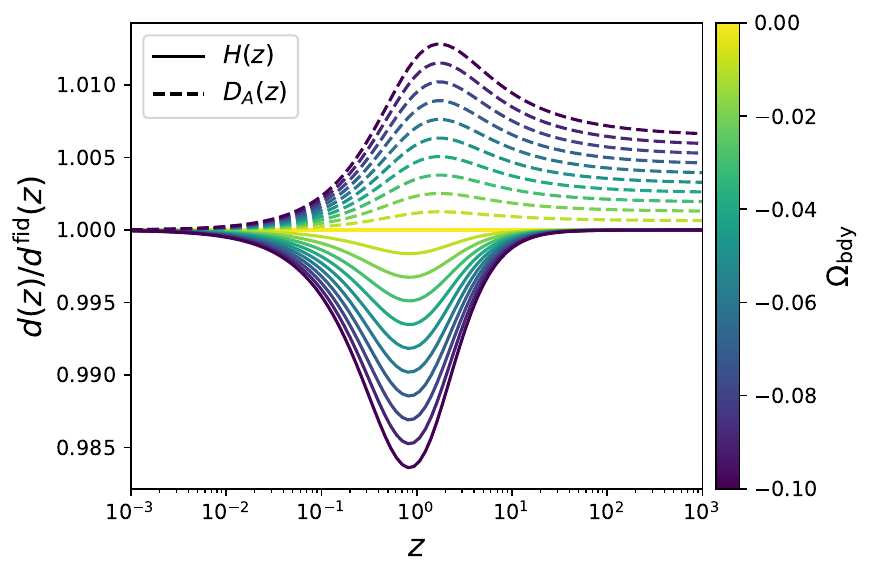}
    \caption{Modification to the Hubble expansion rate (solid) and angular diameter distance (dashed) induced by a collection of topological boundaries with fractional energy density $\Omega_{\rm bdy}$ today. The colored lines show the effect of varying $\Omega_{\rm bdy}$, fixing all other parameters to the DESI+Planck best-fit cosmology except for $\Omega_{\Lambda} = \Omega_\Lambda^{\rm fid}-\Omega_X$ to ensure a flat Universe.  The presence of boundaries reduces the Hubble rate at intermediate redshifts, leading to an increased angular diameter distance for sources at $z\gtrsim 1$.}
    \label{fig: distance-evoln}
\end{figure}

\noindent In the regimes discussed above, the principal effect of the boundaries is to modify the expansion history of the Universe through a negative contribution to the Friedmann equations. Due to the atypical $1/a$ time-dependence its effect is suppressed at recombination (as for dark energy) but becomes more important at late-times, dominating for $(\Omega_{\rm m}/|\Omega_{\rm bdy}|)^{1/2} \lesssim a\lesssim |\Omega_{\rm bdy}|/\Omega_\Lambda$, where $\Omega\equiv 8\pi G/(3H_0^2)\rho_{0}$ is the fractional energy density today. As demonstrated in Fig.\,\ref{fig: distance-evoln}, a negative value of $\Omega_{\rm bdy}$ leads to a reduced expansion rate at intermediate redshifts, and thus increased cosmological distances\footnote{{ See e.g. \cite{Hogg:1999ad} for an overview of various distances in cosmology.}}
; for example, setting $\Omega_{\rm bdy}=-0.1$ increases the distance to $z=0.53$ (the pivot redshift of DESI DR2 \citep{DESI:2025zgx}) by $0.9\%$ and that to the CMB by $0.7$\%, fixing the curvature to zero.  

Since we assume that the boundaries are homogeneously distributed, they do not contribute to the Einstein equations at leading order, except through changes to the background expansion (i.e.\ we assume $|\delta \rho_{\rm bdy}/\rho_{\rm bdy}|\ll 1$ {smoothed over a scale shorter than the horizon size by a suitable numerical control factor}). 
At high redshifts, such effects are negligibly small, thus we expect the boundaries to minimally impact early-Universe quantities of interest, such as the comoving sound horizon at last scattering $r_s(z_*)$, or the scale of matter-radiation equality, $k_{\rm eq}$. The principal effect of boundaries on the CMB is thus a rescaling of the angular coordinates, with more negative $\Omega_{\rm bdy}$ implying a {smaller} acoustic scale $\theta_*\equiv r_s(z_\star)/D_M(z_*)$, where $D_M(z)$ is the comoving angular diameter distance to last scattering. 
We expect a similar effect for the BAO, which is sensitive to the combinations $D_M(z)/r_d$ and $D_H(z)/r_d$, where $r_d\equiv r_s(z_d)$ is the (comoving) sound horizon at baryon drag
and $D_H(z) \equiv c/H(z)$ is the Hubble distance. Since $r_d$ is essentially fixed, both quantities will increase with negative $\Omega_{\rm bdy}$, particularly around $z\sim 1$ (cf.\,Fig.\,\ref{fig: distance-evoln}).

At later times, the modified expansion history can impact the growth of structure. This can be seen from the linearized evolution equation for dark matter:
\begin{eqnarray}
    \frac{d^2}{da^2}\delta_{\rm m}(a) + \left(\frac{3}{a}+\frac{d}{da}\log H(a)\right)\frac{d}{da}\delta_{\rm m}(a) - \frac{3}{2}\frac{\Omega_m(a)}{a^2}\delta_{\rm m}(a) = 0.
\end{eqnarray}
A negative $\Omega_{\rm bdy}$ leads to an increased $\Omega_{\rm m}(a)\propto 1/H^2(a)$ at intermediate times, and thus a slight increase in clustering, with $\sigma_8$ increasing by $1.0\%$ for $\Omega_{\rm bdy}=-0.1$. Besides altering the three-dimensional galaxy clustering signal, this will enhance the CMB lensing signal, both due to an increase in the matter power spectrum and in the lensing kernel itself (due to the increased distance to last scattering), with a $2.2\%$ excess at $L=50$. Furthermore, the increase in $\Omega_\Lambda$ leads to a larger late-time ISW effect, which leads to a slight increase in the large-scale CMB power spectrum. 

\begin{figure}
    \centering
    \includegraphics[width=\linewidth]{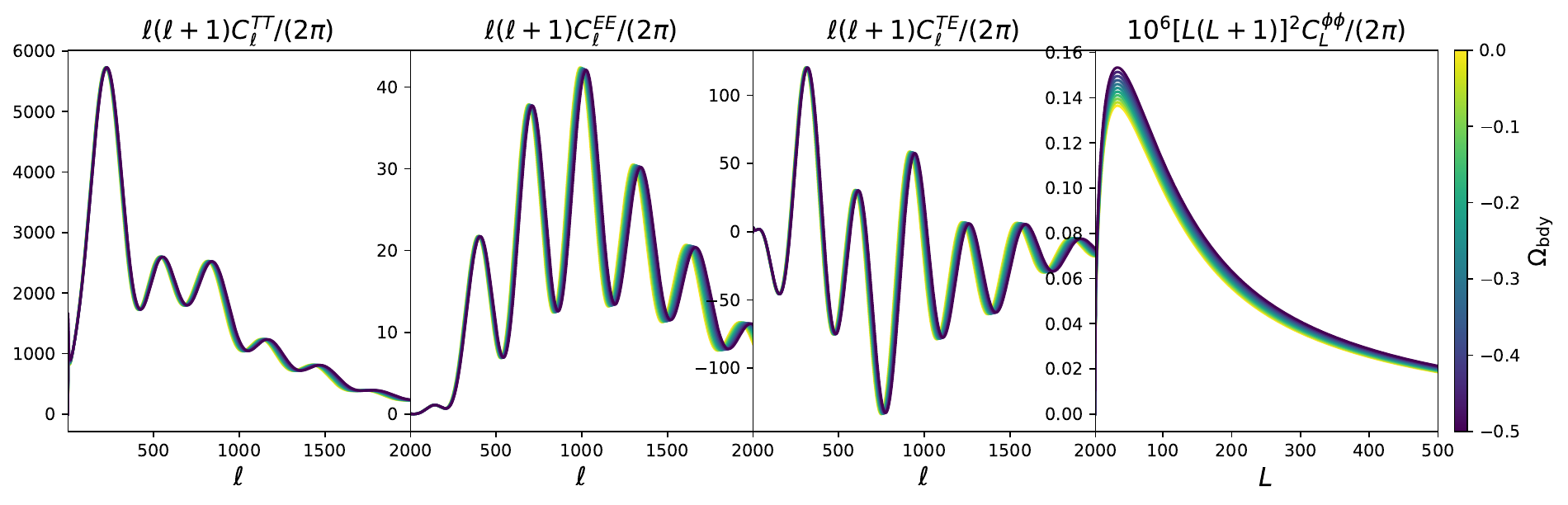}
    \caption{Effect of a set of topological boundaries on the CMB power spectra, fixing all cosmological parameters except $\Omega_\Lambda$. We show results for CMB temperature and polarization auto-spectra (first two panels), cross-spectra (third panel) and the lensing auto-spectrum (fourth panels), with colors indicating the energy density of boundaries. Note that we use a larger range of $\Omega_{\rm bdy}$ than in Fig.\,\ref{fig: distance-evoln} for visibility. As discussed in the text, the main effects of a negative $\Omega_{\rm bdy}$ are an increased distance to last scattering (leading to an angular rescaling), increased lensing power and an increased late-time ISW effect.}
    \label{fig: cmb-effect}
\end{figure}

The combined impact of the above effects on the CMB is shown in Fig.\,\ref{fig: cmb-effect}. This is obtained by a simple modification to the \textsc{camb} code \citep{Lewis:1999bs,camb_code} to add an additional species with density $\rho_{\rm bdy,0}/a$ but without perturbations, i.e.\ $\delta\rho_{\rm bdy}=0$. This demonstrates the above features: an increase to larger $\ell$, an enhancement in the lensing amplitude, and an increased late-time ISW signal (which is difficult to see in the figure). 

In practice, the effects of $\Omega_{\rm bdy}$ can be partially mitigated by a change in the other cosmological parameters. Keeping the acoustic scale fixed (which is exquisitely well measured by Planck \citep{Planck:2018vyg}) requires either increasing $r_s(z_\star)$ or reducing $D_A(z_\star)$; the latter can be achieved by increasing $H_0$ whilst reducing $\Omega_{\rm m}$, such that the physical density $\Omega_{\rm m}h^2$ (which enters the equality scale and $r_s$) remains constant. Similarly, one can suppress the lensing and clustering excess by reducing $A_s$, though this requires a decrease in the reionization optical depth $\tau$ to keep the CMB amplitude constant ($\sim A_se^{-2\tau}$). The change to the ISW cannot be easily hidden however, given its sensitivity to the unusual time dependence of the boundaries. Furthermore, combining the CMB with the BAO data from DESI breaks certain degeneracies (for example allowing direct measurement of $\Omega_{\rm m}$ from the redshift dependence of the Hubble parameter), thus it is not immediately clear whether the above scenario is observationally viable.

\subsection{Comparison to Observational Data}
\noindent Following the above discussion, we now test the boundary model in a Bayesian setting, performing a joint analysis of CMB and LSS data. We utilize the following datasets:
\begin{itemize}
    \item \textbf{CMB}: We use the low-$\ell$ temperature and polarization likelihoods from Planck Public Release 3 (PR3) \citep{Planck:2019nip}, combined with the \textsc{camspec} high-$\ell$ likelihoods from the PR4/\textsc{NPIPE} release \citep{Planck:2020olo,Rosenberg:2022sdy}. We additionally consider the ACT-plus-Planck high-$\ell$ likelihood from ACT DR6 \citep{ACT:2025fju}.
    \item \textbf{Lensing}: We utilize two gravitational lensing likelihoods: Planck PR4 \citep{Carron:2022eyg} and its combination with ACT DR6 \citep{ACT:2023kun}, using the default analysis settings in each case.
    \item \textbf{BAO}: We use the BAO measurements from DESI DR2 \citep{DESI:2025zgx}, including the BGS, LRG, ELG, QSO and Ly-$\alpha$ subsamples.
\end{itemize}
Our default dataset combines the low- and high-$\ell$ Planck likelihoods with Planck PR4 lensing and, optionally, the DESI BAO measurements. To analyze the boundary model discussed above, we perform a Markov Chain Monte Carlo analysis using the \textsc{cobaya} software \citep{Torrado:2020dgo}, asserting convergence when the Gelman-Rubin diagnostic has dropped below $0.01$ for all parameters. For sampling, we utilize a uniform prior on $\Omega_{\rm bdy}\in[-0.5,0.5]$, though additionally consider imposing the constraint $\Omega_{\rm bdy}\leq 0$.

\begin{figure}
    \centering
    \includegraphics[width=0.9\linewidth]{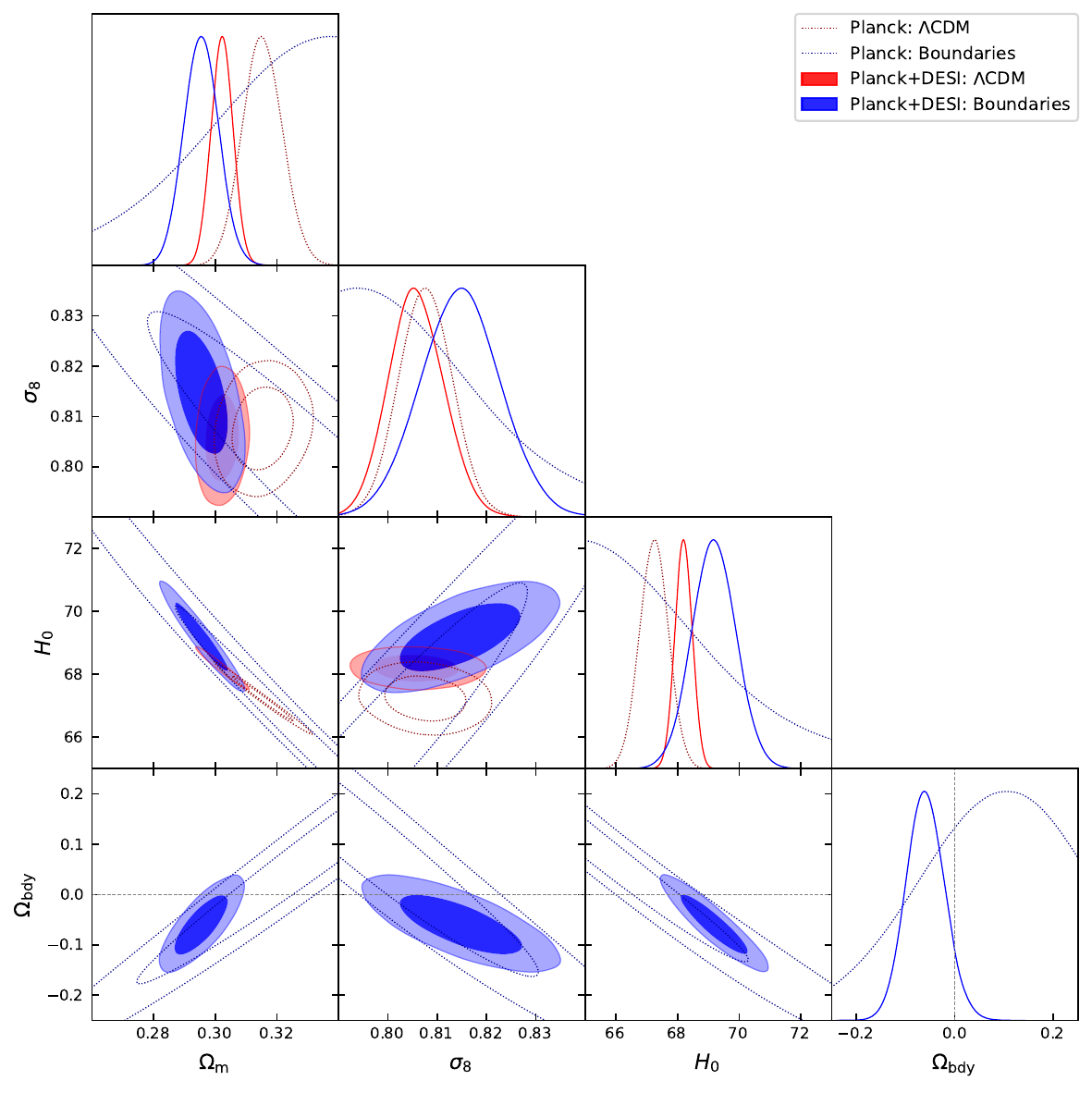}
    \caption{Constraints on cosmological boundaries from CMB and LSS datasets. We show the cosmological parameter posteriors from a likelihood analysis involving \textit{Planck} CMB data (dashed, including lensing) and its combination with DESI BAO data (solid). The red contours indicate $\Lambda$CDM constraints, whilst the blue include a novel species with fractional density $\Omega_{\rm bdy}/a$. Whilst the CMB data alone is unable to break the geometric degeneracies induced by the boundaries, we find a weak preference for negative $\Omega_{\rm bdy}$ (with best-fit $\Omega_{\rm bdy}=-0.059\pm0.039$) when including BAO, as well as a slight shift to larger $\sigma_8$ and smaller $\Omega_m$. The fit to the BAO data are visualized in Fig.\,\ref{fig: bao}, and additional analyses are shown in Fig.\,\ref{fig: corner-alt}\,\&\,\ref{fig: mnu}.}
    \label{fig: corner}
\end{figure}

In Fig.\,\ref{fig: corner} we show the main results of this section: joint constraints on $\Lambda$CDM parameters and the $\Omega_{\rm bdy}$ extension. When considering the CMB data alone, we find weak constraints on the boundaries with $\Omega_{\rm bdy}=0.10\pm0.17$. Notably, $\Omega_{\rm bdy}$ is strongly degenerate with $\Omega_{\rm m}$, $H_0$ and $\sigma_8$; the first two occur due to the lack of redshift information (since we measure only a single physical distance), and the second arises because $\sigma_8$ is an integrated quantity that cannot be directly measured from the CMB (which is sensitive to $A_se^{-2\tau}$), except through lensing.

The inclusion of BAO information leads to considerable degeneracy breaking; a similar effect occurs when one measures the dark energy equation of state from the CMB \citep[e.g.,][]{Planck:2018vyg}. In this case, the posterior on $\Omega_{\rm bdy}$ lies $1.8\sigma$ below zero: $\Omega_{\rm bdy}=-0.059\pm0.039$ or $\Omega_{\rm bdy}=-0.065\pm 0.034$ if we impose the prior $\Omega_{\rm bdy}\leq 0$. These correspond to the 95\% lower bounds 
$\Omega_{\rm bdy}\geq-0.255$ and $\geq -0.123$ respectively. As predicted above, the negative value of $\Omega_{\rm bdy}$ is associated with a shift to smaller $\Omega_{\rm m}$, larger $H_0$ and larger $\sigma_8$; whilst the shift in $H_0$ is towards the values implied by local distance ladder measurements, that in $\Omega_{\rm m}$ is somewhat discrepant with the values preferred by supernovae datasets \citep[e.g.,][]{DESI:2025zgx}.

\begin{figure}
    \centering
    \includegraphics[width=\linewidth]{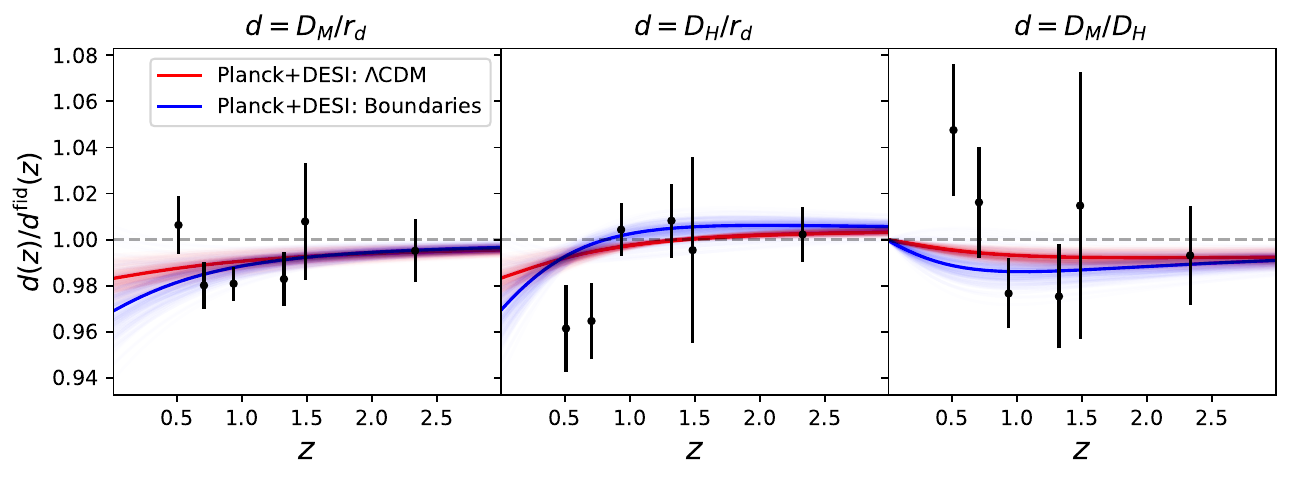}
    \caption{BAO observables in the $\Lambda$CDM model (red) and that with cosmological boundaries (blue), scaling according to the \textit{Planck} best-fit prediction. The points represent the DESI DR2 observations \citep{DESI:2025zpo,DESI:2025zgx} and the lines indicate  draws from the posterior distribution, with the best-fit cosmology shown as a solid line. $D_M$ is the transverse comoving distance (equal to $1+z$ times the angular diameter distance), $r_d$ is the comoving sound horizon at baryon drag and $D_H\equiv c/H$ is the Hubble distance. We omit the BGS sample from the figure, which constrains only the combination $D_M^2D_H$. The boundary model provides a somewhat improved fit to the CMB+BAO dataset by modifying the growth history at intermediate times.}
    \label{fig: bao}
\end{figure}

The analysis suggests a weak preference for our boundary scenario with $\Omega_{\rm bdy}<0$. To understand this it is useful to consider the effects of boundaries on the BAO observables, which are plotted in Fig.\,\ref{fig: bao}. As discussed in \citep{DESI:2025zgx} (and numerous other works), the combination of DESI and Planck prefers a cosmology featuring slightly reduced distances at late-times, which can be well-fit by certain phenomenological dark energy models \citep{DESI:2025wyn,DESI:2025fii}, albeit at the expense of two additional free parameters and an abandonment of the null energy condition.\footnote{ This discussion of NEC violation stimulated our exploration in this work of quantum field theory as a physical source of negative energy (which can become competitive when combined with nontrivial topology).} Whilst the best-fit $\Lambda$CDM model (the red line in Fig.\,\ref{fig: bao}) struggles to fit the BAO dataset, the ansatz of this work (the blue line) offers some additional flexibility due to the new late-time component. Comparing the $\chi^2$ values of the best-fit boundary and $\Lambda$CDM models, we find a weak preference for the former with $\Delta\chi^2=-2.8$ or $-3.0$ when imposing $\Omega_{\rm bdy}\leq 0$ (with one additional degree of freedom). Interestingly, this is set primarily by the change to the high-$\ell$ CMB, with both models obtaining similar fits to the BAO data. This highlights the great constraining power of the CMB; parameter combinations that would improve the BAO fit are themselves tightly constrained from the CMB.

\begin{figure}[ht]
    \centering
    \begin{minipage}[t]{0.47\linewidth}
        \vspace{0pt}
        \centering
        \includegraphics[width=0.9\linewidth]{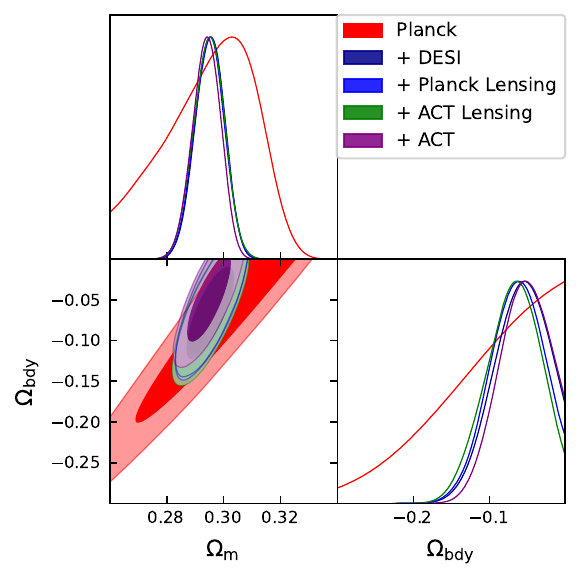}
        \caption{Additional constraints on the cosmological boundary scenario shown in Fig.\,\ref{fig: corner}. We show results from \textit{Planck} alone (red, as before), incorporating DESI BAO without CMB lensing (dark blue), then adding the \textit{Planck} lensing power spectrum (blue, as before), the ACT lensing power spectrum (green), and the full ACT data-vector (purple). We find similar constraints on $\Omega_{\rm bdy}$ in all cases.}
        \label{fig: corner-alt}
    \end{minipage}
    \hfill
    \begin{minipage}[t]{0.45\linewidth}
        \vspace{0pt}
        \centering
        \includegraphics[width=0.9\linewidth]{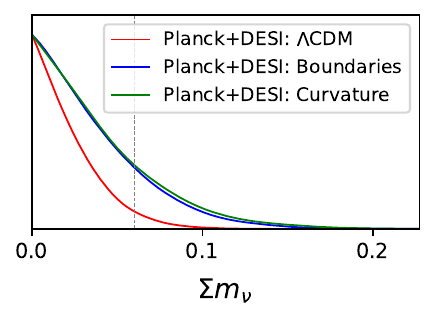}
        \caption{Constraints on the neutrino mass sum (in $\mathrm{eV}$) from the CMB and BAO datasets. In the $\Lambda$CDM model (red), we find a strong constraint with $\sum m_\nu\leq 0.061\mathrm{eV}$ at $95\%$ CL in mild tension with the normal hierarchy prediction; this relaxes to $\sum m_\nu\leq 0.098\mathrm{eV}$ when adding negative-energy boundaries. This is similar to adding negative curvature, which finds $\sum m_\nu.\leq 0.107\mathrm{eV}$ \citep[e.g.,][]{Chen:2025mlf}.}
        \label{fig: mnu}
    \end{minipage}
\end{figure}

In Fig.\,\ref{fig: corner-alt}, we explore the dependence of the constraints on the choice of CMB likelihood, restricting to the $\Omega_{\rm bdy}\leq 0$ model, which had the best $\chi^2$ in our fiducial analysis. Regardless of the choice of lensing likelihood (none, Planck, or Planck-plus-ACT), we find similar constraints on both $\Omega_{\rm m}$ and $\Omega_{\rm bdy}$, with the most negative obtained using Planck-plus-ACT lensing with $\Omega_{\rm bdy}=-0.070\pm0.034$. Switching the primary CMB likelihood leads to a slight increase in the boundary density to $-0.056\pm 0.031$; this model is favored over $\Lambda$CDM at just $\Delta\chi^2 = -1.2$. The relative insensitivity of our constraints to both the lensing and high-$\ell$ CMB likelihoods implies that critical information is contained within large-scale modes -- this agrees with the earlier discussion highlighting the importance of the ISW effect. Indeed, the best-fit boundary model yields an acoustic scale within $0.01\%$ of the $\Lambda$CDM value in our baseline analysis, but features up to $7\%$ larger power in $C_\ell^{TT}$ at $\ell < 30$  {(opposite the direction suggested by large-scale CMB anomalies \citep[e.g.,][]{Jones:2023ncn})}. 


Our data can also place constraints on the sum of the neutrino masses, $\sum m_\nu$ (which was fixed to $0.06\mathrm{eV}$ in our previous analyses). As discussed in \citep[e.g.,][]{Green:2024xbb,DESI:2025ejh,Loverde:2024nfi}, $\Lambda$CDM analyses of Planck and DESI data prefer a low value of $\sum m_\nu$ in some tension with the minimal value $0.06\mathrm{eV}$ expected from solar oscillation experiments \citep[e.g.,][]{Lesgourgues:2006nd}. Indeed, allowing for a (formal) extension to $\sum m_\nu<0$ gives an improved fit to the data, which could indicate a variety of novel physical effects \citep{Craig:2024tky}. In Fig.\,\ref{fig: mnu}, we display our constraints on the neutrino mass from the fiducial Planck-plus-DESI analysis, assuming both the $\Lambda$CDM model and our boundary hypothesis. The addition of free $\Omega_{\rm bdy}\leq 0$ significantly relaxes the constraints: we find $\sum m_\nu \leq 0.098\,\mathrm{eV}$ at 95\% CL, which can be compared to the $\Lambda$CDM bound $\sum m_\nu \leq 0.061\,\mathrm{eV}$. Our value exhibits no tension with the minimal mass, and essentially `trades-off' a negative neutrino mass with a negative boundary density (which is well-motivated physically).

It is instructive to compare our results to those obtained when freeing the spatial curvature, $\Omega_{k}$. As shown in \citep{Chen:2025mlf} (see also \citep{DESI:2025zgx}), the joint Planck-plus-DESI dataset prefers a mildly negative spatial curvature with $\Omega_k = 0.0023\pm0.0011$ (recalling that $\Omega_k \propto -k$); this model yields a significantly enhanced fit to the data with a best-fit $\Delta\chi^2$ of $4.7$ with our data-cuts. Moreover, the bound on the neutrino mass is consistent with that obtained from boundaries: $\sum m_\nu \leq 0.107\,\mathrm{eV}$. Our proposal is similar in spirit to that of curvature: we introduce an additional component which modifies the physical distance to the CMB and the BAO, whilst leaving the early Universe physics roughly invariant.\footnote{From the Friedmann equation \eqref{eq-friedmann-intro}, the two proposals may appear discrepant, since $\Omega_k>0$ leads to an \textit{increase} in $H(z)$ at late-time unlike the decrease expected from $\Omega_{\rm bdy}<0$. This effect is dwarfed by the transformation from comoving to physical distances however.} Indeed, this is the basis of most models invoked to reduce the DESI discrepancy, including time-varying dark energy \citep{DESI:2025wyn,DESI:2025fii,DESI:2024kob}.

\section{Discussion}\label{sec-discussion}

{In this work we introduced a novel type of parameter beyond the cosmological Standard Model. This incorporates generic topology (including boundaries of spacetime), at which quantum fields contribute a UV sensitive boundary layer of vacuum energy which can be consistently negative.  We focused on the effect on the evolution history of a scenario with fixed comoving size boundaries},  {which was shown to be a physically viable regime, and does not appreciably impact the physics of recombination or structure formation.} 

 {When comparing the model to the latest cosmological datasets, we find a slight preference, with a best-fit energy density around $-7\%$ today. This is strongest when combining DESI BAO with the Planck CMB signal (with a maximal $\Delta\chi^2$ of $-3.0$), and is primarily limited by the statistical uncertainties on the expansion history and large-scale CMB modes, through the ISW effect. Furthermore, this scenario relaxes the bound on neutrino masses, analogously to dynamical dark energy and negative curvature. To constrain this model more strongly, one could add supernovae datasets (as in \citep{DESI:2025zgx}), which calibrate the growth-history at low redshifts. Given the variation between different datasets and recent concerns regarding systematic calibration \citep[e.g.,][]{Popovic:2025glk}, we have not attempted such a combination in this work.}

 {It would be interesting to further study the small-scale phenomenology induced by topological boundaries. Though this work focused on regimes in which the main impact of the boundaries was to modify the background expansion, additional effects are possible. In particular, one expected localized distortions of the CMB and LSS around these boundaries (depending on their size and abundance). For suitable boundary conditions, we expect geodesic reflection from the excised regions, which could source atypical signatures in the CMB analogous to strong lensing and the effects of massive inflationary particles \citep{Philcox:2024jpd,Munchmeyer:2019wlh,Flauger:2016idt}. Moreover, we expect non-Gaussian signatures in large-scale structure, on scales of order the boundary size. This rich landscape will be discussed in future work.}


{Finally, it would be interesting to investigate the phenomenological implications of cosmological timelike boundaries for the Standard Model (of particle physics). Certain features of the Standard Model and its extensions (such as chiral fermions, the QCD $\theta$ angle and axion dynamics) can be sensitive to topological properties of spacetime. The presence of such boundaries could thus give rise to novel cosmological signatures, offering new avenues for testing both particle physics and spacetime topology.}

\section*{Acknowledgements}

We would like to thank Colin Hill 
and Albert Law for early collaboration on related problems.  We are also grateful to  Daniel Green and Peter Graham for useful insights on the theoretical status of recent measurements and  to Michael Anderson, Dio Anninos, Eoin Colgain, Don Marolf, Diego Mazzitelli, Jorge Santos, Dick Bond, David Spergel, Glenn Starkman, David Tong, and Zach Weiner for very helpful comments. GT is supported by CONICET (PIP grant 11220200101008CO), CNEA, Universidad Nacional de Cuyo, and by a Simons targeted grant to Instituto Balseiro. OHEP is a Junior Fellow of the Simons Society of Fellows.  The work of ES is supported by a Simons Investigator award and National Science Foundation grant PHY-2310429.   

\appendix

\section{Energy of walls from zero point energies}\label{app:zeropoint}

The goal of this appendix is to review the calculation of the renormalized energy in the presence of walls for a free massless scalar with different boundary conditions. For this, it is useful to calculate the energy by summing zero-point energy contributions for normal modes in the presence of walls, and subtract the zero-point energy from the vacuum in the absence of walls:
\be\label{eq:Eren1}
\langle E_{\rm ren}^{\rm mode}\rangle = \int_0^{M_{\rm UV}} \,d\omega\,\frac{\omega}{2} [N(\omega)-N_0(\omega)]\,.
\ee
Here $M_{\rm UV}$ is a high-energy frequency cutoff.
For a massless scalar field, $N(\omega) d\omega$ is the number of modes
\be
-\bar \nabla^2\,u_\omega(x) = \omega^2 u_\omega(x)
\ee
with frequency in the range $(\omega, \omega+d\omega)$ in the presence of the wall, and $N_0$ is the density in the absence of the wall.

We use heat kernel methods. Let us define the trace of the heat kernel
\be
K(s) = \Tr(e^{-s D})\;,\;D \equiv -\bar \nabla^2\,.
\ee
Denoting the eigenvalues of $D$ by $\lambda=\omega^2$, we can as is standard write this trace in terms of the density of states
\be
K(s) = \int_0^\infty\,d\lambda\,N(\lambda) e^{-s \lambda}\,.
\ee
Note that the density of states for $\omega$ is related to that for $\lambda$ by changing variables in this expression,
\be\label{eq:Nf}
N(\omega) =2 \omega N(\lambda)\,.
\ee
The density of states can be obtained from the trace of the heat kernel by performing an inverse Laplace transform,
\be\label{eq:Nlambda}
N(\lambda) =\frac{1}{2\pi i} \int_C ds\,e^{s \lambda} K(s)\,,
\ee
where $C$ runs between $-i \infty$ and $i \infty$ (plus a possible shift of the real part of $s$ to avoid singularities of $K(s)$).

We want to obtain the contributions to the renormalized energy (\ref{eq:Eren1}) that scale with positive powers of $M_{UV}$. Those are determined by the behavior of the density of states with $\omega \sim M_{UV}$. At large frequency, the integral (\ref{eq:Nlambda}) is dominated by small $s$.
The small $s$ expansion of the heat kernel has the form \cite{Vassilevich:2003xt}
\be\label{eq:Kexpansion}
K(s) = \sum_k s^{(k-n)/2} a_k(D,B)
\ee
where $n$ is the dimension for the manifold in which the operator $D$ is defined ($n=3$ for us), and $B$ denotes the dependence on the boundary condition. {The heat kernel coefficients can be obtained via different approaches, such as solving the DeWitt equation for $\partial_s K$ in an expansion for small $s$, or by employing scale variations and recursive relations \cite{Vassilevich:2003xt}.}

In particular, for Dirichlet, Neumann or Robin boundary conditions, 
\be\label{eq:bcs}
B^- \phi = \phi \Big|_{\partial M}\;\textrm{(Dir)}\;,\;B^+ \phi= (\nabla_n \phi + S \phi)\Big|_{\partial M} \;\textrm{(Robin)}\,.
\ee
the first coefficients were reported in \cite{Vassilevich:2003xt} in formulas (5.16)--(5.19): 
\bea\label{eq:ai}
a_0(D,B^\pm)&=& \frac{1}{(4\pi)^{n/2}} \int_M d^nx\,\sqrt{g} \nonumber\\
a_1(D,B^\pm)&=& \frac{1}{(4\pi)^{(n-1)/2}} \int_{\partial M} d^{n-1}x\,\sqrt{h} b_1^\pm\\
a_2(D,B^\pm)&=& \frac{1}{6(4\pi)^{n/2}}  \left[\int_M d^nx\,\sqrt{g} R+\int_{\partial M} d^{n-1}x\,\sqrt{h} \left(b_2^\pm K + b_4^+ S \right) \right]\,. \nonumber
\eea
Let us explain the different contributions here. $M$ represents the bulk spatial manifold, of dimension $n$, where the operator $D=-g^{ij} \nabla_i \nabla_j$ is defined.\footnote{Additional contributions to $D$ can be easily incorporated as further discussed in \cite{Vassilevich:2003xt}.} It has metric $g$ and Ricci scalar $R$. Its boundary is denoted by $\partial M$, with induced metric $h$; $K$ is the trace of the extrinsic curvature of the boundary. The remaining numerical constants are $b_1^\pm=\pm 1/4$, $b_2^\pm=2$, and $b_4^+=12$.

The inverse Laplace transform (\ref{eq:Nlambda}) for the first three terms in (\ref{eq:Kexpansion}) gives
\be
N(\lambda) =\frac{2\lambda^{1/2}}{\pi^{1/2}} a_0(D,B^\pm)+a_1(D,B^\pm)+ \frac{1}{(\pi \lambda)^{1/2}} a_2(D,B^\pm)+ \ldots
\ee
For the calculation of (\ref{eq:Eren1}) we need $N(\omega)-N_0(\omega)$, which translates into a difference $\Delta a_i$ for the heat kernel coefficients with and without boundary. All the bulk integrals of (\ref{eq:ai}) cancel out in $\Delta a_i$, and only the boundary contributions remain. In particular, $\Delta a_0=0$, $\Delta a_1=a_1$ is proportional to the area of the wall, and $\Delta a_2$ obtains contributions from the integral of the extrinsic curvature and of the parameter $S$ in the Robin condition. Plugging $\Delta N(\omega)=2\omega \Delta N(\lambda)$ into the renormalized energy, we obtain
\be
\langle E_{\rm ren}^{\rm mode}\rangle =\frac{1}{3} \Delta a_1\,M_{\rm UV}^3+\frac{1}{2\pi^{1/2}}\Delta a_2\,M_{\rm UV}^2 + \ldots
\ee

From $\Delta a_1$, we see that for a boundary of size $r_0$ (e.g. the radius of the sphere), Robin and Dirichlet give opposite sign contributions 
\be
\langle E_{\rm ren}^{\rm mode}\rangle \sim \pm r_0^2 M_{\rm UV}^3\,.
\ee
On the other hand, from $\Delta a_2$, they both contribute with the same sign,
\be
\langle E_{\rm ren}^{\rm mode}\rangle \sim  M_{\rm UV}^2 \left(\textrm{sign}(K) r_0+ S r_0^2\right)\,.
\ee
If we are surrounded by a spherical boundary, namely $M$ is given by $0<r<r_0$, the extrinsic curvature is positive and the contribution from this term is $\langle E_{\rm ren}^{\rm mode}\rangle>0$. If, on the other hand, we are outside the wall (i.e. $M$ is given by $r>r_0$), then the extrinsic curvature is negative, and  hence from this term we also obtain a negative contribution to $\langle E_{\rm ren}^{\rm mode}\rangle$. Furthermore, an interesting property of the Robin case is that it provides a low energy parameter $S$ (of dimension 1/length) that can be used to change the sign of the contribution to the wall energy.

{  Similar considerations apply to other types of fields.  One interesting note is that similarly to gravity, in the case of Dirichlet conditions for Yang-Mills fields the symmetry reverts to a global symmetry at the boundary.  It would be interesting to flesh out the implications of this for Standard model gauge and global symmetries and the Fermions in chiral representations.}
{ There is a large body of work on boundary conditions for chiral fermions in the context of confinement and chiral symmetry breaking, starting from the MIT bag model; see e.g. \cite{Johnson:1975zp} for an early review, and \cite{Ivanov:2021yms} for a more modern presentation with an emphasis on heat kernel results. It would be interesting to analyze the dynamics of the Standard Model fermions in the presence of cosmological timelike boundaries of the kind we have considered in the present work.\footnote{The situation for vector-like fermions has been investigated for instance in \cite{BRANSON1992249}. It involves ``mixed boundary conditions'' that relate different spinor components via Dirichlet or Robin type conditions. These boundary conditions lead to a similar expansion to the coefficients $\Delta a_i$ above, and again there are low energy adjustable parameters like $S$ above that can be used to obtain a desired sign for the zero point energy.}}

\bibliographystyle{JHEP}
\bibliography{refs.bib}
\end{document}